\documentclass[runningheads]{llncs}

\usepackage[utf8]{inputenc}
\usepackage{amsmath,amsfonts,amssymb}
\usepackage{pifont}
\usepackage{dsfont}
\usepackage{authblk}
\usepackage{fullpage}
\usepackage{times}
\usepackage{hyperref}
\usepackage{microtype}
\usepackage{color}
\usepackage{cleveref}
\usepackage[colorinlistoftodos]{todonotes}
\usepackage{comment}
\usepackage[ruled, linesnumbered,vlined]{algorithm2e}
\usepackage{algpseudocode}
\usepackage{subcaption}

\newcommand{\cmark}{\ding{51}}
\newcommand{\xmark}{\ding{55}}

\newcommand{\R}{\mathbb{R}}

\newcommand{\sys}{\textsc{Hide \& Seek}\xspace}

\usepackage[]{todonotes}

\newcommand{\vf}{\mathbf{f}}

\newcommand{\set}[1]{\{{#1}\}}

\newcommand{\len}[1]{\lvert{#1}\rvert}

\begin{document}

\title{\sys: Privacy-Preserving Rebalancing on Payment Channel Networks}

\author{Zeta Avarikioti\inst{1} \and
Krzysztof Pietrzak\inst{1}\thanks{Supported by the Vienna Cybersecurity and Privacy Research Center (ViSP), funded by the Vienna business agency (Wirtschaftsagentur),  2020-2023} \and
Iosif Salem\inst{2} \and
Stefan Schmid\inst{2}\thanks{Supported partially by the Austrian Science Fund (FWF) project "Design Framework for Self-Driving Networks" (ADVISE), I 4800-N, 2020-2023 and Vienna Cybersecurity and Privacy Research Center (ViSP), funded by the Vienna business agency (Wirtschaftsagentur),  2020-2023} \and
Samarth Tiwari\inst{3}\thanks{Supported partially by ERC Starting Grant QIP--805241 and the Vienna Cybersecurity and Privacy Research Center (ViSP), funded by the Vienna business agency (Wirtschaftsagentur),  2020-2023} \and
Michelle Yeo\inst{1}}

\authorrunning{Z. Avarikioti et al.}

\institute{IST Austria \\
\email{\{zetavar, krzysztof.pietrzak, michelle.yeo\}@ist.ac.at} \and
Faculty of Computer Science, University of Vienna, Austria\\
\email{\{iosif.salem, stefan\_schmid\}@univie.ac.at} \and
Centrum Wiskunde \& Informatica, Amsterdam, The Netherlands\\
\email{samarth.tiwari@cwi.nl}
}

\maketitle

\begin{abstract}

Payment channels effectively move the transaction load off-chain thereby successfully addressing the inherent scalability problem most cryptocurrencies face. 
A major drawback of payment channels is the need to ``top up'' funds on-chain when a channel is depleted. 
Rebalancing was proposed to alleviate this issue, where parties with depleting channels move their funds along a cycle to replenish their channels off-chain. Protocols for rebalancing so far either introduce local solutions or compromise privacy.

In this work, we present an opt-in rebalancing protocol that is both private and globally optimal, meaning our protocol maximizes the total amount of rebalanced funds. 
We study rebalancing from the framework of linear programming.
To obtain full privacy guarantees, we leverage multi-party computation in solving the linear program, which is executed by selected participants to maintain efficiency.
Finally, we efficiently decompose the rebalancing solution into incentive-compatible cycles which conserve user balances when executed atomically.

\end{abstract}

\section{Introduction}
Cryptocurrencies are increasingly growing as an alternative payment method. By replacing a central trusted authority (e.g., a bank) with a decentralised ledger, i.e., a blockchain, mutually distrusting users now have the means to achieve consensus over transactions. However, achieving consensus on the blockchain is notoriously inefficient. Bitcoin, for instance, can only support at most 7 transactions per second on average~\cite{poon2015lightning}. This severely limits the scalability of blockchain solutions to every day life situations.

Payment channel networks (PCNs) aim to increase the efficiency and scalability of blockchains while maintaining the benefits of security and decentralisation.
PCNs operate on top of blockchains introducing Layer 2 -- the blockchain itself being Layer~1.
As the name suggests, a PCN consists of several payment channels between pairs of users who wish to transact with each other. Users connected indirectly through a path of channels may route transactions through the network.
To open a payment channel, two users create a funding transaction where they lock funds on-chain only to be used in this payment channel.
Thereafter, each transaction on the payment channel is simply an exchange of a signed message that depicts the current balances between the two users; so it does not involve the blockchain at all.
This can go on indefinitely until the users go back to the blockchain to close the channel. 
The process of closing a channel  consists  of one on-chain transaction optimistically, while in worst case of a small constant number of transactions (e.g., in Lightning closing a payment channel  costs  at most two on-chain transactions).
Thus, with at most three blockchain transactions, any pair of users can in theory make an arbitrary number of costless transactions with each other.

A major drawback of payment channels is that users cannot simply ``top up" their balance in the channel off-chain once it is depleted. 
Instead, they have to go on-chain to refund the payment channel. A solution to extend the lifetime of payment channels is \textit{rebalancing}, which updates payment channels with the crucial condition that the overall balance of each node is unchanged. Although it is not possible to shift funds from one payment channel to another off-chain, the effect of rebalancing is precisely that: funds from well-funded payment channels transfer to depleted ones.

There are two predominant approaches to rebalancing. The first involves a local search of rebalancing cycles (i.e., transactions of a fixed amount that begin and end with the same user) initiated by a single user. This is the current rebalancing approach in the Lightning Network \cite{lightningrebalancinggitrepo}. The second approach (introduced in ~\cite{khalil2017revive}) is global instead of local: nodes looking to rebalance specify a maximum amount of rebalancing flow along each of their channels, where the rebalancing transactions are determined by a global evaluation of the state of the network.

\begin{figure}[htb!]
    \centering
    \includegraphics[scale=0.7]{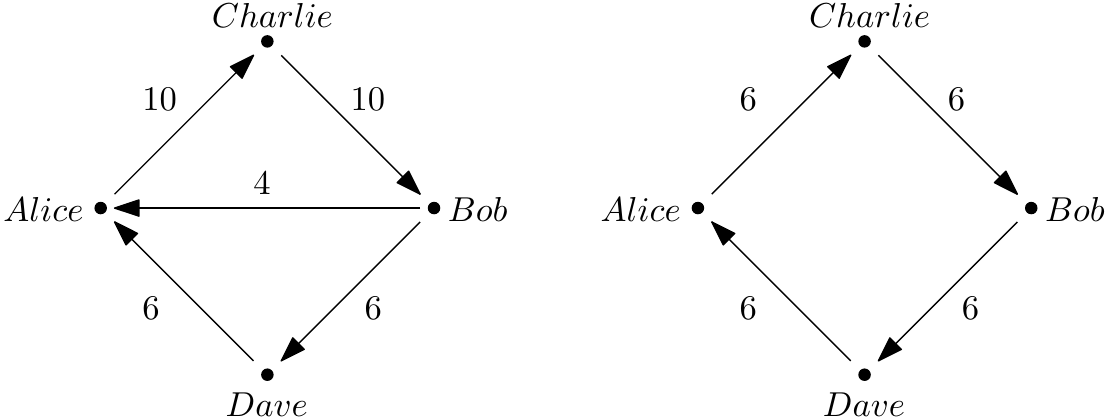}
    \caption{The cancelling out effect. The weighted, directed edges on the graph on the left specifies the maximum coins a user can forward along the direction of the edge. The graph on the right shows the maximum rebalancing cycle Charlie can achieve using the cycle finding approach.}
    \label{fig:cancellingout}
\end{figure}

A drawback of single-user based cycle finding is it overlooks other rebalancing requests across the network, leading to \textit{local solutions}.  Figure~\ref{fig:cancellingout} illustrates one such consequence which we call the ``cancelling out'' effect. Suppose a user Charlie wants to move 10 coins from his channel with Bob to his channel with Alice. If Charlie utilises the cycle finding approach, he will only manage to rebalance 6 coins as depicted in the graph on the right. The channel between Bob and Alice would be ignored because of the lack of sufficient balance on Bob's end. In a globally optimal solution, however, the entire rebalancing in the graph on the left can be executed, as it takes into account that transactions in both directions can be above the capacity of a channel, as long as they ``cancel out" and the resulting transaction is within the capacity.

Furthermore, users must check if the other users on the cycle are willing to forward the rebalancing transaction amount, even after finding rebalancing cycles. This could lead to a prolonged and laborious search for cycles with willing participants. Lastly, this approach requires users to have global knowledge of the network topology which can be unrealistic in terms of storage as the size of the network increases. 

The second approach does not suffer from local limitations such as the canceling out effect, and theoretically achieves the global optimal rebalancing. Revive~\cite{khalil2017revive} implemented this method by assigning a random delegate, either a trusted external third party or someone from the set of participants, to receive channel constraints and solve a linear program that models rebalancing. This is a serious \textit{privacy loophole}, since the delegate now has information on the concerned payment channels.
Moreover, the delegate has control over the rebalancing output; for instance, the delegate may compute the rebalancing transactions in a malicious or suboptimal way, favouring some transactions over others.
Although the authors proposed a method for any participant to challenge the rebalancing transactions, the process is lengthy and requires giving the challenger access to the balances of all participants. 

In this work, we present \sys, the first opt-in rebalancing protocol that is both \textit{private} and achieves a \textit{globally optimal rebalancing}.
Each party that is interested in rebalancing specifies the maximum amount to be forwarded in each of the party's channel.
We employ selected delegates that receive the maximum amounts per channel, calculate and share with each party the exact amount to be moved in each channel.
We formulate our problem as a linear program and set our objective function to maximize the total amount of funds to be rebalanced in the network.
Our protocol does not involve transaction fees.

On the other hand, we leverage multi-party computation to obtain a \textit{fully private solution}. 
Specifically, the participants in \sys only learn the information they would have learned if a trusted third party computed the optimal rebalancing and returned to each participant the amount to be moved along each of their channel. No sensitive information such as the channel balances is leaked.

Finally, we guarantee the rebalancing can be securely and efficiently executed. We propose a simple way to decompose the optimal rebalancing circulation into a set of transaction cycles. 
As a result, the transactions of each cycle are easy to execute atomically using HTLCs. We note that atomicity is limited to each rebalancing cycle, therefore increasing the protocol's robustness; 
any cycle can be executed successfully regardless of the success of other cycles.

We highlight the advantages of our approach in Table~\ref{tab:summary}.

\begin{table*}
\begin{center}
\begin{tabular}{ |c|c|c|c|c|c| } 
\hline
 & Private & Globally optimal & Opt-in & Network locality\\
\hline
Cycle finding solution & \cmark & \xmark & \xmark & \xmark\\ 
\hline
Revive  & \xmark & \cmark & \cmark & \cmark \\ 
\hline
\textbf{Our solution}& \cmark & \cmark & \cmark & \cmark\\
\hline
\end{tabular}
\end{center}
\caption{Summary of main approaches for rebalancing. Private solutions are solutions that do not leak balance information. Globally optimal refers to the optimality of the rebalancing solution. Opt-in refers to solutions where willing users choose to participate in the protocol and non-willing users are not involved at all. Network locality refers to solutions that only require local knowledge of the PCN.}
\label{tab:summary}
\vspace{-20pt}
\end{table*}

\subsubsection{Our contributions.}

We introduce \sys, the first opt-in privacy-preserving and globally optimal rebalancing protocol that can be implemented in a secure and efficient manner. We acknowledge and discuss its limitations in terms of efficiency. We suggest several practical speed ups for the deployment of our solution, and outline possible extensions.

\section{Preliminaries}

\subsection{Payment channels networks} 
Users $u, v$ can open a payment channel between each other by locked some of their funds to be used only in this channel: if $u$ locks $a$ units and $v$ locks $b$ units, the state of the channel from $u$ to $v$ is modeled as a real number $\text{balance}(u,v) \in [-b,a],$ initialized as $0$. The capacity of the channel refers to the sum $a+b$ of these funds. Once the channel is created, both users can send each other money by updating the channel balances in favour of the other party, as long as the state remains within the interval $[-b,a]$. 

Users who are not directly connected by a channel in a PCN can still transact with each other if they are connected by a path of payment channels. The users along the transaction path which are not the sender or receiver typically charge a fee for forwarding the transaction that depends on the transaction amount. For a transaction to be successful, the sender has to first send enough money to cover both the desired payment amount and all the fees charged by each user on the payment path. That is, suppose user $s$ wants to send $x$ coins to user $r$ along a payment path $p = \{ (s, u_1), ..., (u_k, r) \}$. Then $s$ must send $x+ \sum_{i=1}^k\text{fee}(u_i)$. Secondly, the balance of each user along the path must be large enough to forward the payment amount together with fees. Then for each user $u_i$ on $p$, $\text{balance}(u_i, u_{i+1}) \geq x + \sum_{j=i+1}^k \text{fee}(u_j)$. Although the channel capacities are typically public information, the individual balances on each end are private; so senders typically have to try different payment paths until one of them succeeds.

A desired guarantee for payment routing through a path in a PCN is atomicity, i.e., for all users along the path, either all of them update their balances or none of the balances in the path get updated. This is enforced in the Lightning Network using HTLCs \cite{lnpaper}. An HTLC ($HTLC(u,v,x,h,t)$) is a smart contract between any two users $u$ and $v$ that locks some amount of coins $x$ using a hash output $h$ and a timelock $t$. To get the locked funds, $v$ has to produce the preimage $r$ to the hash $h = H(r)$ within time $t$, upon which the locked funds will be released to $v$. If $v$ cannot do so within the time limit, $u$ can claim the locked funds. Payment path atomicity is enforced using HTLCs for each channel on the path with the same hash value (determined using a secret chosen by the receiver on the path), but with decreasing timelock values from sender to receiver to guarantee the security of funds.

\subsection{Network flows}
Consider a directed graph $G = (V, E)$ and the associated $\len{E}$-dimensional Euclidean space of non-negative flow along each edge. A circulation is a flow $\vf = (f(u,v))_{(u,v) \in E}$ such that the net flow through each vertex is zero: $\sum\limits_{v \in V} f(u,v) = \sum\limits_{v\in V} f(v,u), \forall u \in V$. Two circulations $\vf_1, \vf_2$ can be added to get yet another circulation: $\vf_1 + \vf_2 = (f_1(u,v) + f_2(u,v))_{(u,v) \in E}$. A cycle is a sequence of vertices $v_1, v_2 \ldots v_k$ such that $(v_i, v_{i+1}) \in E, \forall 1 \leq i \leq k-1$ and $(v_k, v_1) \in E$ as well. We may equivalently refer to this cycle as $(e_1, e_2 \ldots e_k)$ where $e_i = (v_i, v_{i+1}), \forall 1 \leq i \leq k-1$ and $e_k = (v_k, v_1)$. We call $k$ the length of this cycle. A cycle flow $\vf$ of weight $w$ on cycle $C$ is a circulation where $f(e) = w, \forall e \in C$ and $f(e) = 0$ otherwise.

A standard result of network flow theory is that any circulation may be expressed as a sum of at most $\len{E}$ cycles. We refer the reader to the textbook of Ahuja, Magnanti and Orlin(~\cite{Ahuja1993}) for a detailed treatment. 

\section{Protocol Overview and Model}

\subsection{System model}

\paragraph{Payment network topology.}
We model the PCN as a graph $\Tilde{G} = (\Tilde{V}, \Tilde{E})$, with a vertex for each node and an edge between $u$ and $v$ if there is a payment channel between them.
Let $V \subset \Tilde{V}$ be the users in the PCN that are interested in rebalancing and let $G=(V,E)$ be the subgraph of $\Tilde{G}$ induced by $V$.
We denote $\len{V} = n, \len{E} = m$. We assume each user $u$ has only local knowledge of the PCN topology, i.e., only knows the capacities and balances on the edges incident to $u$.

\paragraph{Cryptographic assumptions.}
We assume the existence of secure communications channels, hash functions and signatures. We follow \cite{DamgardN03} and assume the concept of an arithmetic black box for MPC $\mathcal{F}_{ABB}$, in particular with functionalities like secret sharing, storage, retrieval, addition, multiplication, and comparisons.

\paragraph{Blockchain \& network model.}
We assume a synchronous network, i.e., there is known bounded message delay. We further assume the underlying blockchain satisfies persistence and liveness as defined in~\cite{garay2015bitcoin}.




\subsection{Protocol overview}
In a nutshell, our proposed protocol \sys consists of two phases: an exploration phase and an execution phase.
Firstly, the goal of the exploration phase is to discover rebalancing cycles privately and efficiently. 
Then, the goal of the execution phase is to guarantee that the rebalancing transactions are executed in a secure manner. 
At the same time we want to maximise the efficacy of our protocol, that is, we want as many rebalancing cycles to go through as possible.  

\paragraph{Exploration phase.}
The exploration phase first formulates the rebalancing problem as a linear program. Then we randomly select $k$ delegates out of the participants to perform an MPC protocol to jointly solve the linear program. Next, any set of participants that wish to participate in the rebalancing protocol prepares the shared inputs to the delegates. The output of the exploration phase is a rebalancing circulation. 

\paragraph{Execution phase.}
We first efficiently decompose the rebalancing circulation output of the exploration phase into a set of cycle flows. These cycle flows have the property that they are sign-consistent, i.e. they are consistent with the direction of the flows in the rebalancing circulation. This makes executing these cycles incentive-compatible, as no user would have to execute transactions which violate their specified rebalancing capacity and direction along channels. Once this is done, we enforce atomicity of these cycles by creating an HTLC for each cycle which ensures either transactions along the the entire cycle goes through or none at all.

\subsection{Desired properties \& threat model}
\label{sec:desired}
In general, we assume a computationally bounded  adversary, i.e., runs in probabilistic polynomial time.
The properties \sys should guarantee are the following:
\begin{enumerate}
    \item \textbf{Balance conservation (security):} The total balance of each node, which is the sum of the node's balances on each incident channel, must remain the same before and after \sys, even when \textit{all other participants are corrupted} by the adversary.
    \item \textbf{Privacy:} The information revealed during \sys should be not exceed the minimum required to execute rebalancing: (a)~the participants must only learn the transaction amounts for each of their payment channels; (b)~the delegates of MPC should not be able to determine private financial information of the participants. Both (a) and (b) should hold as long as \textit{one of the delegates is not corrupted}.
    \item \textbf{Optimality (completeness):} Assuming \textit{every participant is honest}, the result should be optimal in that no other rebalancing yields greater total change over all payment channels.
\end{enumerate}

\section{The \sys Protocol}
\subsection{Exploration phase}
\subsubsection{Linear programming for rebalancing.} 
The practical problem of rebalancing has many facets, including keeping participants' financial information private and facilitating coordination. We overlook these considerations momentarily to present the underlying optimization problem of rebalancing. 

For a payment channel between $(u,v)$, the users would like to move the state $\text{balance}(u,v)$ towards a desired state $\text{balance}^*(u,v)$.
If $\text{balance}^*(u,v) > \text{balance}(u,v)$ then rebalancing would involve $u$ transferring funds to $v$, and we model this as a directed edge from $u$ to $v$ with capacity $m(u,v) := \text{balance}^*(u,v) - \text{balance}(u,v)$. If $\text{balance}^*(u,v) < \text{balance}(u,v)$ then there is a directed edge from $v$ to $u$ with capacity $m(v,u) := \text{balance}(u,v) - \text{balance}^* (u,v)$. Thus the graph $G$ is transformed into a directed weighted graph.

The capacities $m(u,v)$ represent the most flow that can occur through each channel during rebalancing. If $m(u,v)=0$ the edge from $u$ to $v$ is either non-existent or equivalently, a zero-capacity edge. We also enforce that if $m(u,v) > 0$, then necessarily $m(v,u)=0$.

Let us denote a potential rebalancing by $\vf \in \R^{\len{E}}$ on this directed graph, where  $f(u, v)$ denote the flow from $u$ to $v$. Since rebalancing should not result in a net financial gain or loss for any participant, we require $\vf$ to be a circulation. Recall that it means the net flow through each vertex is zero:
$$\sum\limits_{v: (u, v) \in E} f(u, v) = \sum\limits_{v: (v, u) \in E} f(v, u).$$

Not only must the flows be non-negative, but they must also satisfy the capacity constraints as specified by participants:
$$0 \leq f(u, v) \leq m(u, v).$$

Thus, the set of valid rebalancings is a polytope in $m$-dimensional Euclidean space defined by $n + 2m$ linear constraints: $n$ zero flow constraints for each vertex and $m$ pairs of flow capacity constraints for each edge.

We wish to compute a rebalancing that maximizes the linear objective $ \sum\limits_{(u, v) \in E} f(u, v)$. We call the linear program so specified the rebalancing problem. This choice of objective function amounts to maximizing the total change in each payment channel's balance towards its desired state.

\subsubsection{Solving the rebalancing problem.}
One can apply any linear programming algorithm of preference to solve the rebalancing, such as any from the family of simplex methods. In fact, the rebalancing problem can be reduced to the min-cost flow problem, a specialization of linear programming which can be solved more easily. For instance, the min-cost flow problem admits a strongly polynomial algorithm, meanwhile the corresponding question for linear programming is a major open problem in the field. 

Appendix~\ref{app:mincost} illustrates how the rebalancing problem is equivalent to a min-cost flow problem with the same number of vertices and edges. Henceforth, we refer to the rebalancing problem as a min-cost flow problem.

\subsubsection{Delegate selection and multi-party computation.}
Delegate selection can be done using a simple version of cryptographic sortion as in \cite{gilad2017algorand}. Each of the $k$ delegates involved in the MPC gets $n+2m$ inputs which are shares of each of the $n$ participant's zero flow constraints and the $2m$ rebalancing capacity constraint along the $m$ directed edges (for each edge we have two constraints: one which specifies the maximum rebalancing flow in one direction, and another which specifies the flow has to be 0 in the other direction). The objective function is also shared and given as an input to the delegates. The delegates jointly compute the optimal solution to the rebalancing LP problem and each delegate outputs a share of the final flow on each edge at the end of the protocol.

\subsection{Execution phase}

\subsubsection{Cycle decomposition.}
The exploration phase concludes with a solution to the rebalancing linear program obtained through multi-party computation. This solution $\vf^*$ is in fact encoded as shared secrets, and, as observed in \cite{Toft09} (relevant passage), one can process the solution further before returning to individual participants. Instead of directly sending each $f^*(u,v)$ to $u$, we decompose the circulation into a sum of cycle flows. This makes the execution of rebalancing  via HTLCs easier; instead of the entire network committing their funds to a large atomic rebalancing transaction, each cycle only requires coordination between nodes constituting the cycle. 

\begin{figure}[t]
    \centering
    \includegraphics[scale=0.9]{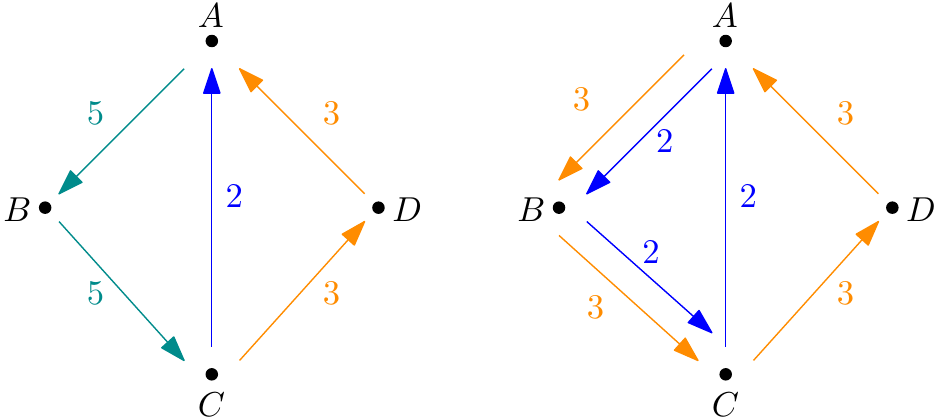}
    \caption{The graph on the left depicts a circulation. The weight of each edge is the transaction amount to send along the edge. The cycles in the graph on the right is a sign consistent decomposition of the the circulation.}
    \label{fig:cycledecomp}
\end{figure}

As mentioned earlier, 
each circulation can be expressed as a sum of cycle flows. We briefly describe a standard algorithm to compute this decomposition efficiently. Algorithm~\ref{algo:cycledecomp} uses depth-first search as a subroutine to detect cycles and then induce cycle flows on them. Figure \ref{fig:cycledecomp} depicts a circulation and its decomposition into cycle flows.

\begin{algorithm}[t]
	\SetKwInOut{Input}{input}\SetKwInOut{Output}{output}
	\DontPrintSemicolon
	\Input{ Circulation $\vf$ on directed graph $G = (V,E)$ }
	\Output{ A set of cycle flows $\mathcal{S}$ that sum to $\vf$}
	initialize $i=1$\\
	initialize $R \longleftarrow \set{e \in E: f(e) \neq 0}$ set of active edges\\
	\While{$R \neq \emptyset$}{
	    pick an edge $e_1 \in R$\\
	    run depth first search to find a cycle $C_i = (e_1, e_2, \ldots e_k)$ in $R$\\
	    $w_i \longleftarrow \min f(e), e \in C_i$\\
	    initialize $\vf_i \longleftarrow 0$\\
	    \For{$e \in C_i$}{
	        $f_i(e) = w_i$\\
	        $f(e) \longleftarrow f(e) - f_i(e) $\\
	        \If{$f(e) = 0 $}{
	            delete $e$ from $R$
	        }
	    }
	    $i \longleftarrow i + 1$
	}
	\Return{$\mathcal{S} = \set{\vf_1, \vf_2 \ldots \vf_i}$}
	\caption{Depth-first Search Cycle Decomposition}
	\label{algo:cycledecomp}
\end{algorithm}

\subsubsection {HTLC commitments per cycle.}
Given such a decomposition, we need to enforce atomicity of each cycle flow by creating an HTLC for each cycle $c$ in the set. This can be done by first selecting a user in each cycle at random to initiate the cycle. This user has to choose a random secret $r_c$ from some domain $\mathcal{X}$ and create a hash of the secret $h_c = H(r_c)$. The timelock for the initiator of the cycle and the next user is set equal to the length of the cycle. The transaction amount to send along each cycle is the weight of the cycle $w_c$. Every subsequent user in the cycle decrements the timelock value by 1 and looks up the next user in the cycle they should create an HTLC with (determined by the vertex order in the cycle). They then create an HTLC with that user with the decremented timelock value (lines 5-7 in Algorithm \ref{algo:HTLCcycles}).

\begin{algorithm}[t]
	\SetKwInOut{Input}{input}\SetKwInOut{Output}{output}
	\DontPrintSemicolon
	\Input{ $\mathcal{S}$ set of directed cycles}
	\For{$c \in \mathcal{S}$}{
	    select starting user $u_c$ at random from users in $c$\\
	    timelock $t_c \longleftarrow len(c)$ \\
	    $u_c$ chooses random secret $r_c$ and creates hash $h_c = H(r_c)$\\
	    \For{$e_c = (u,v) \in c \text{ starting from } u_c$}{
	        $u$ creates $HTLC(u, v, w_c, h_c, t_c)$\\
	        decrement $t_c$ by 1
	    }
	}
	\caption{HTLC creation for cycles}
	\label{algo:HTLCcycles}
\end{algorithm}

Finally we note that Algorithm \ref{algo:cycledecomp} and Algorithm \ref{algo:HTLCcycles} can be computed privately using MPC. To prevent any two users on a cycle $c$ from sharing their hash $h_c$ with each other and thus finding out they are in the same cycle, one can use MAPPCN \cite{tripathy2020mappcn} to preserve user anonymity.

\section{Analysis}
\subsubsection{Desired Properties.}
An execution of \sys satisfies the desired properties as stated in Section \ref{sec:desired}.
%
Let us study each of the properties in order.

\paragraph{Balance Conservation.} Suppose there is a node $v$ that enjoys net financial gain through the execution of \sys under a malicious adversary. \sys specifies a set of cycle flows that the nodes may execute, and $v$ must have participated in some subset of these. Note that by atomicity of cycle flows ensured through HTLCs, it is not possible for a cycle to be executed partially (even when parties act maliciously). If $v$'s balance increased, that means there must be at least one cycle flow with net positive flow through $v$. But this contradicts the definition of cycle flows, since they must satisfy zero flow through each node:
$$\sum\limits_{(u,v) \in E} f(u,v) = \sum\limits_{(v,u)\in E} f(v,u) \quad \forall v \in V.$$

\paragraph{Privacy.} The sensitive data used in the exploratory phase of \sys remains private as long as at least one delegate of the MPC is honest (inherited by the MPC). In the execution phase, users do not know the other users in their cycle except their predecessors and successors in their cycles as we use MAPPCN to preserve user anonymity.


\paragraph{Optimality.} Assuming the delegates compute the solution correctly, the circulation returned by the min-cost flow algorithm maximizes the total flow through each edge. Under the same assumption, the cycle decomposition algorithm would result in an equivalent (and thus also optimal) set of cycle flows. 


\subsubsection{Efficiency.}
We break the analysis of the efficiency of \sys into three parts: (1) solving the rebalancing problem, (2) cycle decomposition, and (3) MPC.

\paragraph{Solving the rebalancing problem.}
Solving the underlying min-cost flow problem is the most computationally intensive aspect of \sys. Fortunately we can leverage the vast body of algorithms for this problem being asymptotically optimal in different parameter regimes. The complexity of these algorithms is analyzed in terms of $n, m$, the largest capacity $U$ and the largest cost $W$ of an edge. We may presently ignore the term $W$ as each edge has identical cost $1$. For the parameter regime of rebalancing, we recommend the double scaling algorithm of Ahuja, Goldberg, Orlin and Tarjan with computes the optimal solution in time $O(nm \log nW \log \log U)$ (\cite{Ahuja92}).

An alternative is to use a network simplex algorithm. This family of algorithms are excellent in practice, although theoretical analysis of their effectiveness is an active area of research in optimization. Simplex algorithms are also incredibly simple, and for this reason have they been recommended in Toft's framework for privately solving linear programs(\cite{Toft09} despite the somewhat poorer theoretical guarantees. We recommend the network simplex algorithm of Orlin (\cite{Orlin96}) for the rebalancing problem, which terminates in at most $O(nm \log n)$ pivots. Generally, the amortized cost per pivot is $O(n)$, but Orlin presents a modification with total runtime $O(nm \log n \log n W)$.

\paragraph{Cycle decomposition.}
If the rebalancing circulation obtained by solving the min-cost flow contains $n'$ vertices and $m'$ edges, then the cycle decomposition algorithm as detailed in Algorithm~\ref{algo:cycledecomp} terminates in $O(n'm')$ time, which is $O(nm)$ at worst. Every loop iteration removes at least one edge, and each iteration visits at most $n'$ vertices before finding a cycle. The pre-processing of $G$ to obtain the subgraph induced by the circulation takes $O(n+m)$ time. 

Also note that the timelocks used in the execution of a cycle flow are bounded by the length of the cycle.

\paragraph{MPC.}
Although MPC implementations of optimization algorithms incur a penalty in speed, there are multiple methods to speed up the implementation of \sys:

Firstly, the rebalancing problem, much like many other min-cost flow problems, satisfies the Hoffman-Gale conditions: the optimal solution, along with the vertices of the polytope, is guaranteed to be integral. This means the MPC can be performed over faster integer arithmetic rather than slower floating point arithmetic. 

\sys can be implemented even faster by reducing the number of bits per variable. This quantity is governed by the maximum capacity per edge as well as the granularity of rebalancing, so that the number of bits required depends on the specific cryptocurrency. For instance, an implementation of \sys for Bitcoin with just $20$ bits per variable may restrict all quantities to multiples of $2^{10} =1024$ satoshis up to $2^{30}$ satoshis which is approximately $10$ bitcoins.

The number of delegates chosen to compute the MPC also contributes to the communication cost during rounds, and here we note that \sys does not place any limitations on this number. In fact, it can be as low as two delegates as long as one of them is honest. 

Finally, the efficiency of our protocol inherently depends on the MPC primitives used. This is a wide and active area of research, with a lot of new developments in making efficient MPC primitives \cite{CramerFIK03,Aly2015,Octavian2010,crypto-2020-30428}.

\section{Limitations and Extensions}
\label{sec:limitations}
In this section, we identify the limitations of our protocol and discuss possible extensions.

\subsubsection{Rational participants.}
Participants in financial networks such as PCNs typically act selfishly, aiming to increase their financial gain. As a result, an interesting future study is the security of our scheme under rational participants. 
In the execution phase, the cycle decomposition ensures that participants always gain from executing a cycle because the cycles are sign-consistent. Nevertheless, HTLCs have been proven vulnerable to attacks where participants collude and act for-profit~\cite{malavolta2019anonymous}.
Regarding the exploration phase, it has been shown that when participants are rational (with respect to privacy) MPC is possible using randomized mechanisms with constant expected running time~\cite{abraham2006distributed}. 

\subsubsection{Weighted LP.}
The linear program of the rebalancing problem currently maximizes the total flow through each edge in the network. This is but an approximation of the practical objective, since in practice, flows through distinct edges are not necessarily equally important.

A more accurate model of rebalancing involved modifying the objective function from $\sum_{(u,v) \in E} f(u, v)$ to $\sum_{(u,v) \in E} w(u, v) f(u, v)$ for non-negative integral weights $w(u,v)$ supplied by $u$ via secret sharing. Let $W$ be the maximum possible weight that participants may specify. 

This slight modification greatly enlarges the expressive power of participants, as they can now provide local preferences of one cycle over another. For instance, a user $u$ with one outgoing edge $e_0$ and three incoming edges $e_1, e_2, e_3$ wishes to rebalance $e_0$ desperately. $u$ considers rebalancing along $e_1$ favorable but not urgent, is indifferent to rebalancing along $e_2$, and does not permit any flow through $e_3$. Knowing that outgoing flow through $e_0$ must be balanced by equal incoming flow, $u$ may assign a weight $w(e_2) = 0$ to allow for flow through $e_2$ and then $e_0$ in order to rebalance $e_0$. This edge preference can be expressed by weights:
$$w(e_0) = W, 
\qquad w(e_1) = 1, \qquad w(e_2) = 0,$$
and by not including $e_3$ in the protocol at all.

The desired properties of \sys continue to hold after this modification. In terms of efficiency, the double scaling algorithm that we use runs in $O(nm \log nW \log \log U)$ time rather than $O(nm \log n \log \log U)$~\cite{Ahuja92}.

The major drawback of this modification is game theoretic:
although incorporating preferences is straightforward when users faithfully follow the protocol, it breaks under the assumption of rational participants. In particular, misreporting the weight of every edge as the maximum $W$ is a dominant strategy, since that assigns the highest possible weight to every \textit{cycle} that a user is part of. This reduces this modification to the original case of maximizing $\sum\limits_{(u,v) \in E} f(u, v)$. An improved design of this mechanism, such as a clever budgeting of weights, could circumvent this problem, manage individual users' incentives, and let the weighted LP extension be used practically.

\subsubsection{Optimality with corrupted participants.}
Participants' sensitive financial data, such as existence of a payment channel and its capacity for rebalancing, is not verified in the protocol, nor does our threat model consider falsification of this data with respect to optimality.

Unfortunately, this lack of verification can prevent any rebalancing to occur: an adversary with knowledge of the payment channel network can falsify edge data so that each cycle passes through one of their edges. The adversary can then refuse to participate in the execution phase and prevent others from rebalancing, even when cycle flows between honest parties exist.

To defend against such adversary, we propose that parties submit zero knowledge proof of validity along with their edge constraint data. Although one cannot force participants to participate in rebalancing cycles, this modification certainly increases the success rate of rebalancing cycles in \sys even under an active adversary.

\section{Related Work}
\paragraph{Rebalancing PCNs.}
There are several payment channel primitives proposed in literature~\cite{spilman2013channels,poon2015lightning,decker2015fast,avarikioti2019brick,avarikioti2020cerberus,miller2017sprites}. Regardless of the primitive, a challenge all PCNs share is how to route transactions in the PCN while maintaining balanced channels for as long as possible.
Classic routing studies in PCNs like SilentWhispers~\cite{malavolta2017silentwhispers},  SpeedyMurmurs~\cite{roos2017settling}, 
and others ~\cite{prihodko2016flare} ignore that channels may be slowly depleting.
A promising approach to avoid channel depletion and prolong the network availability for transaction routing is to maintain balanced  channels or occasionally perform rebalancing.
But transaction routing is a challenging task on its own because the channel balances remain secret for privacy purposes~~\cite{khalil2017revive,giulia,CoinExpress}, let alone avoiding channel depletion on-top.

Khalil and Gervais introduce the first channel rebalancing protocol, called Revive~\cite{khalil2017revive}. They formulate the problem as an LP, similarly to our work. Then, a delegate is elected to solve the LP and return the solution to the rebalancing participants. Although our work lies close to Revive, it also differs in several aspects.
First, Revive considers rebalancing as an LP as well, but \sys employs faster and more specific min-cost flow algorithms. Second, Revive relies on a single delegate to compute the optimal rebalancing which leaks private information about balances to the delegate. In contrast, \sys uses MPC to achieve full privacy guarantees. Since \sys uses MPC, the speeds of the two protocols cannot be compared. We nevertheless expect Revive to also benefit from using our min-cost flow framework. Finally, atomic execution of the rebalancing transactions in Revive requires the transaction language of the underlying blockchain to be Turing-complete, and thus it is not suitable for Bitcoin. \sys avoids this issue by first decomposing the optimal rebalancing into cycles, and then executing these cycles atomically using HTLCs. The cycle decomposition in \sys also ensures that, as long as the channel is not part of all cycles, some rebalancing can still occur if individual HTLCs fail on a channel.

From a practical perspective, rebalancing in the Lightning Network currently utilises a brute force search for rebalancing cycles with sufficient capacity. An automated approach for doing so using the imbalance measure was proposed by \cite{pickhardt2020imbalance}. Unlike \sys, these methods do not leverage other rebalancing requests to find the globally optimal rebalancing. These methods also require nodes to have global knowledge of the network whereas nodes in \sys only need to have local knowledge of the PCN.


Recently some works introduce routing protocols that attempt to maintain balanced channels.
In particular, Spider~\cite{giulia} is a payment routing algorithm that maximizes the throughput while maintaining the original channel balances, without providing rebalancing however. 
Li et al. \cite{Li2020} propose to extend the lifetime of payments channels by estimating payment demand, and using this estimate to decide on the initial balance of channels. 
Engelshoven and Roos~\cite{engelshoven2020}, on the other hand, leverage routing fees to incentivize the balanced use of payment channels.
All these works are orthogonal and complementary to ours, as we introduce an opt-in rebalancing protocol.

\paragraph{Network flows and MPC.}
The general problem of solving network flow problems via multi-party computation is considered in the comprehensive PhD thesis of Aly~\cite{Aly2015}. Various privacy preserving implementations of combinatorial optimization problems are presented. The author acknowledges that the cost for privacy is very high even for the simplest of problems. Roughly speaking, their MPC implementations must iterate for the theoretical worst-case number of iterations 
to maintain privacy.
For the practical problem of rebalancing though, we do not choose to implement extra iterations. On the other hand, we believe that suboptimal rebalancing is better than no rebalancing, and recommend terminating the min-cost flow solution prematurely if needed. Both scaling algorithms and network simplex algorithms monotonically generate better solutions in each iteration, leaving the participants with a feasible solution if they stop early.

\section{Conclusion and Future Work}
In this work we study the rebalancing problem for PCNs. We present \sys, which is a secure opt-in rebalancing protocol, that is also private and finds the globally-optimal rebalancing. \sys achieves better efficiency by reducing the rebalancing problem to a min-cost flow problem. \sys also achieves better robustness by decomposing the solution into cycles and executing each cycle atomically, as opposed to executing the entire solution atomically. 

An interesting direction for future work is to consider the transaction aggregation problem, which is similar to rebalancing but without the balance conservation property (for instance Alice's balance is not conserved if Alice wants to pay Bob 2 coins for a coffee). The main difficulty with transaction aggregation comes from the constraint that transactions may not be executed partially. In other words, where the optimization underlying rebalancing is a linear program (solvable in polynomial time), the problem underlying transaction aggregation is an integer program (NP-complete in general).

\bibliographystyle{splncs04}
\bibliography{bibfile}

\begin{thebibliography}{10}
\providecommand{\url}[1]{\texttt{#1}}
\providecommand{\urlprefix}{URL }
\providecommand{\doi}[1]{https://doi.org/#1}

\bibitem{lightningrebalancinggitrepo}
Rebalance plugin.
  \url{https://github.com/lightningd/plugins/tree/master/rebalance}

\bibitem{abraham2006distributed}
Abraham, I., Dolev, D., Gonen, R., Halpern, J.Y.: Distributed computing meets
  game theory: robust mechanisms for rational secret sharing and multiparty
  computation. In: PODC (2006), \url{https://doi.org/10.1145/1146381.1146393}

\bibitem{Ahuja1993}
Ahuja, R., Magnanti, T., Orlin, J.: Network flows - theory, algorithms and
  applications (1993)

\bibitem{Ahuja92}
Ahuja, R., Goldberg, A., Orlin, J., Tarjan, R.: Finding minimum-cost flows by
  double scaling. Math. Program.  \textbf{53},  243--266 (02 1992).
  \doi{10.1007/BF01585705}

\bibitem{Aly2015}
Aly, A.: Network flow problems with secure multiparty computation (2015)

\bibitem{avarikioti2019brick}
Avarikioti, Z., Kogias, E.K., Wattenhofer, R., Zindros, D.: Brick: Asynchronous
  incentive-compatible payment channels. In: FC (2021),
  \url{https://fc21.ifca.ai/papers/168.pdf}

\bibitem{avarikioti2020cerberus}
Avarikioti, Z., Litos, O.S.T., Wattenhofer, R.: Cerberus channels:
  Incentivizing watchtowers for bitcoin. In: FC (2020),
  \url{10.1007/978-3-030-51280-4_19}

\bibitem{crypto-2020-30428}
Baum, C., Orsini, E., Scholl, P., Soria-vazquez, E.: Efficient constant-round
  mpc with identifiable abort and public verifiability. Springer-Verlag (2020)

\bibitem{Octavian2010}
Catrina, O., de~Hoogh, S.: Secure multiparty linear programming using
  fixed-point arithmetic. In: Gritzalis, D., Preneel, B., Theoharidou, M.
  (eds.) Computer Security -- ESORICS 2010. pp. 134--150. Springer Berlin
  Heidelberg, Berlin, Heidelberg (2010)

\bibitem{CramerFIK03}
Cramer, R., Fehr, S., Ishai, Y., Kushilevitz, E.: Efficient multi-party
  computation over rings. In: Biham, E. (ed.) Advances in Cryptology -
  {EUROCRYPT} 2003, International Conference on the Theory and Applications of
  Cryptographic Techniques, Warsaw, Poland, May 4-8, 2003, Proceedings. Lecture
  Notes in Computer Science, vol.~2656, pp. 596--613. Springer (2003).
  \doi{10.1007/3-540-39200-9\_37},
  \url{https://doi.org/10.1007/3-540-39200-9\_37}

\bibitem{DamgardN03}
Damg{\aa}rd, I., Nielsen, J.B.: Universally composable efficient multiparty
  computation from threshold homomorphic encryption. In: Boneh, D. (ed.)
  Advances in Cryptology - {CRYPTO} 2003, 23rd Annual International Cryptology
  Conference, Santa Barbara, California, USA, August 17-21, 2003, Proceedings.
  Lecture Notes in Computer Science, vol.~2729, pp. 247--264. Springer (2003).
  \doi{10.1007/978-3-540-45146-4\_15},
  \url{https://doi.org/10.1007/978-3-540-45146-4\_15}

\bibitem{decker2015fast}
Decker, C., Wattenhofer, R.: A fast and scalable payment network with bitcoin
  duplex micropayment channels. In: Stabilization, Safety, and Security of
  Distributed Systems (2015), \url{10.1007/978-3-319-21741-3_1}

\bibitem{engelshoven2020}
van Engelshoven, Y., Roos, S.: The merchant: Avoiding payment channel depletion
  through incentives. CoRR  \textbf{abs/2012.10280} (2020),
  \url{https://arxiv.org/abs/2012.10280}

\bibitem{garay2015bitcoin}
Garay, J., Kiayias, A., Leonardos, N.: The bitcoin backbone protocol: Analysis
  and applications. In: Eurocrypt (2015), \url{10.1007/978-3-662-46803-6_10}

\bibitem{gilad2017algorand}
Gilad, Y., Hemo, R., Micali, S., Vlachos, G., Zeldovich, N.: Algorand: Scaling
  byzantine agreements for cryptocurrencies. In: SOSP (2017),
  \url{10.1145/3132747.3132757}

\bibitem{lnpaper}
Joseph~Poon, T.D.: The bitcoin lightning network: Scalable off-chain instant
  payments. Tech. rep.,
  \url{https://lightning.network/lightning-network-paper.pdf}

\bibitem{khalil2017revive}
Khalil, R., Gervais, A.: Revive: Rebalancing off-blockchain payment networks.
  In: CCS (2017), \url{10.1145/3133956.3134033}

\bibitem{Li2020}
Li, P., Miyazaki, T., Zhou, W.: Secure balance planning of off-blockchain
  payment channel networks. In: IEEE INFOCOM 2020 - IEEE Conference on Computer
  Communications. pp. 1728--1737 (2020).
  \doi{10.1109/INFOCOM41043.2020.9155375}

\bibitem{malavolta2017silentwhispers}
Malavolta, G., Moreno{-}Sanchez, P., Kate, A., Maffei, M.: Silentwhispers:
  Enforcing security and privacy in decentralized credit networks. In: NDSS
  (2017), \url{10.14722/ndss.2017.23448}

\bibitem{malavolta2019anonymous}
Malavolta, G., Moreno{-}Sanchez, P., Schneidewind, C., Kate, A., Maffei, M.:
  Anonymous multi-hop locks for blockchain scalability and interoperability.
  In: NDSS (2019), \url{10.14722/ndss.2019.23330}

\bibitem{miller2017sprites}
Miller, A., Bentov, I., Bakshi, S., Kumaresan, R., McCorry, P.: Sprites and
  state channels: Payment networks that go faster than lightning. In: FC
  (2019), \url{10.1007/978-3-030-32101-7_30}

\bibitem{Orlin96}
Orlin, J.: A polynomial time primal network simplex algorithm for minimum cost
  flows. Math. Prog.  \textbf{78},  109--129 (01 1996).
  \doi{10.1007/BF02614365}

\bibitem{pickhardt2020imbalance}
Pickhardt, R., Nowostawski, M.: Imbalance measure and proactive channel
  rebalancing algorithm for the lightning network. In: {IEEE} International
  Conference on Blockchain and Cryptocurrency, {ICBC} 2020, Toronto, ON,
  Canada, May 2-6, 2020. pp.~1--5. {IEEE} (2020).
  \doi{10.1109/ICBC48266.2020.9169456},
  \url{https://doi.org/10.1109/ICBC48266.2020.9169456}

\bibitem{poon2015lightning}
Poon, J., Dryja, T.: The bitcoin lightning network: Scalable off-chain instant
  payments. \url{https://lightning.network/lightning-network-paper.pdf} (2015)

\bibitem{prihodko2016flare}
Prihodko, P., Zhigulin, S., Sahno, M., Ostrovskiy, A., Osuntokun, O.: Flare: An
  approach to routing in lightning network. \url{shorturl.at/adrHP} (2016)

\bibitem{roos2017settling}
Roos, S., Moreno-Sanchez, P., Kate, A., Goldberg, I.: Settling payments fast
  and private: Efficient decentralized routing for path-based transactions.
  arXiv preprint arXiv:1709.05748  (2017)

\bibitem{giulia}
Sivaraman, V., Venkatakrishnan, S.B., Ruan, K., Negi, P., Yang, L., Mittal, R.,
  Fanti, G., Alizadeh, M.: High throughput cryptocurrency routing in payment
  channel networks. In: 17th $\{$USENIX$\}$ Symposium on Networked Systems
  Design and Implementation ($\{$NSDI$\}$ 20). pp. 777--796 (2020)

\bibitem{spilman2013channels}
Spilman, J.: Anti dos for tx replacement.
  \url{https://lists.linuxfoundation.org/pipermail/bitcoin-dev/2013-April/002433.html},
  accessed: 2020-11-22

\bibitem{Toft09}
Toft, T.: Solving linear programs using multiparty computation. In: Dingledine,
  R., Golle, P. (eds.) Financial Cryptography and Data Security, 13th
  International Conference, {FC} 2009, Accra Beach, Barbados, February 23-26,
  2009. Revised Selected Papers. Lecture Notes in Computer Science, vol.~5628,
  pp. 90--107. Springer (2009). \doi{10.1007/978-3-642-03549-4\_6},
  \url{https://doi.org/10.1007/978-3-642-03549-4\_6}

\bibitem{tripathy2020mappcn}
Tripathy, S., Mohanty, S.K.: Mappcn: Multi-hop anonymous and privacy-preserving
  payment channel network. In: International Conference on Financial
  Cryptography and Data Security. pp. 481--495. Springer (2020)

\bibitem{CoinExpress}
Yu, R., Xue, G., Kilari, V.T., Yang, D., Tang, J.: Coinexpress: A fast payment
  routing mechanism in blockchain-based payment channel networks. In: 2018 27th
  International Conference on Computer Communication and Networks (ICCCN).
  pp.~1--9. IEEE (2018)

\end{thebibliography}

\appendix

\section{Reduction of The Rebalancing Problem to min-cost Flow}
\label{app:mincost}
Recall that the rebalancing problem consists of finding a circulation on a directed graph with maximum flow while also satisfying the capacity constraints. The related well-studied problem of min-cost circulation provides a cost to each edge as well as lower and upper bounds on the flow through each edge. Rebalancing can thus be seen as a circulation problem with negative costs with flow bounds given by $0$ and capacity $m(u,v)$. Below, we provide a short reduction to the more fundamental min-cost flow problem on the same graph.

The reduction is a simple change of variables: define $\vf' \in \R^m$ as $f'(v,u) := m(u,v) - f(u,v)$. Consider the reversed graph $G' = (V, E')$ where all directed edges from $G$ are reversed $E'= \set{e' = (v,u): (u,v) \in E}$. Rebalancing on $G$ is equivalent to a min-cost flow problem on $G'$. 

The constraints $0 \leq f(u,v) \leq m(u,v)$ transform into $0 \leq f'(v,u) \leq m'(v,u) = m(u,v)$. Finally, the zero flow constraints from the rebalancing problem
$$\sum\limits_{(u,v) \in E} f(u,v) - \sum\limits_{(v,u) \in E} f(v,u) = 0$$
transform into
$$\sum\limits_{(v,u) \in E'} m(u,v) - f'(v,u) - \sum\limits_{(u,v) \in E'} m (v, u) - f'(u,v) = 0,$$
or,
$$\sum\limits_{(u,v) \in E'} f'(u,v) - \sum\limits_{(v,u) \in E'} f'(v,u) = \sum\limits_{(u,v) \in E} m(u,v) - \sum\limits_{(v,u) \in E} m(v,u)$$
In other words, the sources and sinks can be defined by whether $\sum\limits_{(u,v) \in E} m(u,v) - \sum\limits_{(v,u) \in E} m(v,u)$ is positive or negative. By a standard technique, we can further reduce the problem to that containing a single source and single sink by appending so-called ``super-source and super-sink" to $G'$.

Finally, we need to specify the cost to complete the problem description: if the objective of rebalancing is to maximise $\sum\limits_{(u,v) \in E} c(u,v)f(u,v)$ then we specify the min-cost flow problem to minimize $\sum\limits_{(u,v) \in E'} c(v,u)f'(u,v)$. In this way, not only are the feasible regions of both problems equivalent by the described change of variables, but so are the optimum solutions.

\end{document}